\documentclass[12pt]{article}
\usepackage{a4}
\usepackage{amsfonts}
\usepackage{amsmath,amssymb}

\begin{document}

\thispagestyle{empty}

\vskip2.0cm
\begin{center}

{\Large\bf Inflation, Dark Energy and Dark Matter \\ 
\vglue.1in in Supergravity}

\vglue.3in
Sergei V. Ketov~${}^{a,b,c}$ 
\vglue.1in

${}^a$~Department of Physics, Tokyo Metropolitan University, \\
1-1 Minami-ohsawa, Hachioji-shi, Tokyo 192-0397, Japan\\
${}^b$~Research School of High-Energy Physics, Tomsk Polytechnic University,\\
2a Lenin Avenue, Tomsk 634028, Russian Federation \\
${}^c$~Kavli Institute for the Physics and Mathematics of the Universe (WPI),
\\The University of Tokyo Institutes for Advanced Study, Kashiwa, Chiba 277-8583, Japan 
\vglue.3in

{\it Talk presented at the 2019 Meeting of the Division of Particles and Fields of the American Physical Society (DPF2019), July 29 - August 2, 2019, Northeastern University, Boston, C1907293}

\end{center}

\vglue.3in

\begin{center}
{\Large\bf Abstract}
\end{center}
\vglue.2in
The Dark Side of the Universe, which includes the cosmological inflation
in the early Universe, the current dark energy and dark matter, can be theoretically
described by supergravity, though it is non-trivial. We recall the arguments pro and contra
supersymmetry and supergravity, and define the viable supergravity models describing 
the Dark Side of the Universe in agreement with all current observations. Our approach to inflation 
is based on the Starobinsky model, the dark energy is identified with the positive cosmological constant 
(de Sitter vacuum), and the dark matter particle is given by the lightest superparticle identified with the
supermassive gravitino. The key role is played by spontaneous supersymmetry breaking.

\section{Introduction}	

The standard approach to the Dark Side of the (current) Universe is based 
on the Cosmological Concordance Model that assumes (i) the dark energy  
described by the cosmological constant and (ii) the dark matter (DM) described
by unknown (electrically neutral and stable) massive particles. In addition,
the standard (single-field) approach to cosmological inflation in the early Universe 
assumes the existence of another scalar field (called inflaton) driving inflation. The
early Universe inflation can be considered as the primordial dark energy because
"dark energy" is merely the substitute for the accelerating expansion of the Universe,
though the underlying physics of inflation and (current) dark energy are different:
one needs an appropriate scalar potential describing inflaton slow roll for inflation, 
whereas  a de Sitter vacuum (the positive cosmological constant) suffices for dark energy.
As regards the cold dark matter particle, one needs to explain its origin and find the reason
for its stability. 

In supergravity theory, getting all those features is highly non-trivial, because
they have to be compatible with local supersymmetry of the acton.  However, this  problem
simultaneously offers the opportunity to severely constrain possible options for embedding
the Cosmological Concordance Model into supergravity and connect it to the Standard Model (SM)
of elementary particles ~\cite{Ketov:2012yz}.

Let us briefly review the existing theoretical motivation for supersymmetry (SUSY) and supergravity.
SUSY is the one of the leading candidates for new physics beyond the SM. Supergravity is the field theory with 
local SUSY that automatically implies the general coordinate invariance. The minimal ($N=1$) supergravity in four
spacetime dimensions is chiral that is necessary for particle phenomenology and CP violation. Supergravity has 
many attractive features:
\begin{itemize}
\item SUSY unifies bosons and fermions;
\item supergravity automatically includes General Relativity  (GR);
\item supergravity is the conservative extension of GR and field theory, which restricts the number of independent parameters (coupling constants);
\item SUSY Grand Unified Theories (super-GUTs)  result in the perfect unification of electro-weak and strong interactions;
\item the spectrum of matter-coupled supergravities with spontaneously broken SUSY has the natural DM candidate given by the Lightest SUSY Particle (LSP), provided that R-parity is conserved;
\item SUSY helps to stabilize the fundamental scales (the hierarchy problem), such as the electro-weak scale and the GUT scale;
\item SUSY leads to cancellation of the quadratic UV-divergences in quantum field theory;
\item  supergravity is the only way to consistently describe coupling of spin 3/2 particles to gravity;
\item it is unknown how many degrees of freedom (d.o.f.) were present during inflation. Supergravity may be the answer;
\item supergravity can be considered as the low-energy effective action of superstrings (quantum gravity).
\end{itemize}

Of course, some of those arguments may not survive in the ultimate theory. It is also worthwhile to recall the
standard arguments {\it against} SUSY and supergravity, and briefly comment on their validity:
\begin{itemize}
\item the masses of bosons and fermions are not equal. This means that SUSY, if exists, has to be spontaneously broken in our Universe;
\item no supersymmeric partners to the SM particles were found at the Large Hadron Collider (LHC) so far. This merely means
that the SUSY scale is considerably higher that 10 TeV;
\item the observed cosmological constant is positive. This means that one has to find the supergravity models admitting de Sitter vacua;
\item the observed CP violation rules out extended SUSY. This means that one should use the minimal (chiral) 
supergravity in four spacetime dimensions for phenomenological purposes. This supergravity may be derivable
from extended supergravities in higher spacetime dimensions;
\item there is the huge matter-over-antimatter abundance in our Universe. First, we may not observe a considerable
part of antimatter in the Universe. Second, the Affleck-Dine baryogenesis mechanism based on SUSY
can accommodate any matter-over-antimatter abundance~\cite{dine};
\item Superstring Landscape has over $10^{500}$ vacua. First, the very idea of Superstring Landscape is
speculative. Second, the superstring theory itself does not have enough  theoretical tools to address the multiverse concept.
\end{itemize}
In summary, SUSY and supergravity are still alive and are not ruled out by observations despite of the current absence of any signs of their presence at the LHC and in the sky.

In this paper we very briefly review the standard approach to cosmology in supergravity and its problems (Sec.~2).
Then we outline our approach and its tools based on the vector supermultiplet unifying inflaton and goldstino (Sec.~3). The  DM in our approach is given by the superheavy gravitino LSP in the context of high-scale SUSY (Sec.~4). In Sec.~5 we outline
the minimalistic approach entirely based on the use of a single massive vector multiplet with the Born-Infeld (BI) structure and
 the alternative Fayet-Iliopoulos (FI) terms. We conclude in Sec.~6 where we also comment on a description of the primordial
 black holes in supergravity.

\newcommand{\be}{\begin{equation}}
\newcommand{\ee}{\end{equation}}
\newcommand{\nbe}{\begin{equation*}}
\newcommand{\nee}{\end{equation*}}

\newcommand{\fr}{\frac}
\newcommand{\lb}{\label}


\def\a{\alpha}
\def\b{\beta}
\def\c{\chi}
\def\d{\delta}
\def\e{\epsilon}
\def\f{\phi}
\def\g{\gamma}
\def\h{\eta}
\def\i{\iota}
\def\j{\psi}
\def\k{\kappa}
\def\l{\lambda}
\def\m{\mu}
\def\n{\nu}
\def\o{\omega}
\def\p{\pi}
\def\q{\theta}
\def\r{\rho}
\def\s{\sigma}
\def\t{\tau}
\def\u{\upsilon}
\def\x{\xi}
\def\z{\zeta}
\def\D{\Delta}
\def\F{\Phi}
\def\G{\Gamma}
\def\J{\Psi}
\def\L{\Lambda}
\def\O{\Omega}
\def\P{\Pi}
\def\Q{\Theta}
\def\S{\Sigma}
\def\U{\Upsilon}
\def\X{\Xi}


\def\ve{\varepsilon}
\def\vf{\varphi}
\def\vr{\varrho}
\def\vs{\varsigma}
\def\vq{\vartheta}


\def\ca{{\cal A}}
\def\cb{{\cal B}}
\def\cc{{\cal C}}
\def\cd{{\cal D}}
\def\ce{{\cal E}}
\def\cf{{\cal F}}
\def\cg{{\cal G}}
\def\ch{{\cal H}}
\def\ci{{\cal I}}
\def\cj{{\cal J}}
\def\ck{{\cal K}}
\def\cl{{\cal L}}
\def\cm{{\cal M}}
\def\cn{{\cal N}}
\def\co{{\cal O}}
\def\cp{{\cal P}}
\def\cq{{\cal Q}}
\def\car{{\cal R}}
\def\cs{{\cal S}}
\def\ct{{\cal T}}
\def\cu{{\cal U}}
\def\cv{{\cal V}}
\def\cw{{\cal W}}
\def\cx{{\cal X}}
\def\cy{{\cal Y}}
\def\cz{{\cal Z}}

\def\fracm#1#2{\hbox{\large{${\frac{{#1}}{{#2}}}$}}}
\def\fracmm#1#2{{{#1}\over{#2}}}

\section{Inflation in gravity and supergravity}

The Starobinsky model of inflation is defined by the action~\cite{Starobinsky:1980te}
 \be \label{star}
S_{\rm Star.} = \fracmm{M^2_{\rm Pl}}{2}\int d^4x\sqrt{-g} \left( R +\fracmm{1}{6m^2}R^2\right)~,
\ee
where we have introduced the reduced Planck mass $M_{\rm Pl}=1/\sqrt{8\p G_{\rm N}}\approx
2.4\times 10^{18}$ GeV, and the scalaron (inflaton) mass $m$ as the parameter. We use the
spacetime signature $(-,+,+,+,)$. 

The (modified) gravity action (1) is known to be classically equivalent to the scalar-tensor gravity or the standard (quintessence) model of the (canonical) scalar field $\varphi$ minimally coupled to Einstein gravity
and having the scalar potential
\begin{equation} \label{starp}
V(\varphi) = \fracmm{3}{4} M^2_{\rm Pl}m^2\left[ 1- \exp\left(-\sqrt{\frac{2}{3}}\varphi/M_{\rm Pl}\right)\right]^2~.
\end{equation}
This scalar potential has the Minkowski vacuum at  $\varphi=0$  and the {\it plateau} of the positive height (related to the inflationary energy density) that gives rise to slow roll of inflaton during the inflationary era.

The Starobinsky model  (\ref{star}) is the excellent model of cosmological inflation, in very good agreement with the Planck data~\cite{planck2018}.  The Planck satellite mission measurements of the Cosmic Microwave Background (CMB) radiation 
give the scalar perturbations tilt as $n_s\approx 1+2\eta_V -6\ve_V = 0.9649\pm 0.0042$ (with 68\% CL) and restrict the tensor-to-scalar ratio as $r\approx 16\ve_V < 0.064$ (with 95\% CL). The Starobinsky inflation yields $r\approx 12/N_e^2\approx 0.004$ and $n_s\approx 1- 2/N_e$, where $N_e$ is the e-foldings number between 50 and 60, with the best fit at $N_e\approx 55$.

The inflationary model (\ref{star}) is truly geometrical because it is entirely based on gravitational interactions. The inflaton mass parameter $m$ is fixed by the observed CMB amplitude (COBE, WMAP) as 
\be \lb{starm}
m\approx 3 \cdot10^{13}~{\rm GeV} \quad {\rm or}\quad \fracmm{m}{M_{\rm Pl}}\approx 1.3\cdot 10^{-5}~.
\ee

As regards the duration of Starobinsky inflation, a numerical analysis yields
\be\lb{sandf}
\sqrt{\fracmm{2}{3}} \varphi_{\rm ini.}/M_{\rm Pl} \approx \ln\left( \fracmm{4}{3}N_e\right) \approx 5.5~,
\quad \sqrt{\fracmm{2}{3}} \varphi_{\rm end}/M_{\rm Pl} \approx \ln\left[ \fracmm{2}{11}(4+3\sqrt{3})\right]\approx 0.5~~,
\ee
where we have used the best fit $N_e\approx 55$.

As regards supergravity extensions of inflationary models, there are two obvious possibilities for local supersymmetrization: either
taking the modified gravity models or the quintessence models. However, first, one has to embed inflaton field into a supermultiplet, when one wants linearly realized local SUSY. There are two natural options for assigning the inflaton supermultiplet to matter (with the highest spin under or equal to $1$): either a chiral multiplet or a (single) vector multiplet.
 
The standard approach to inflation and DE in supergravity is based on the use of chiral superfields with the maximal spin $1/2$, which requires complexification of inflaton by adding a physical pseudo-scalar (called sinflaton). Two chiral superfields are generically needed: the one including inflaton and another one including goldstino, because inflation implies spontaneous SUSY breaking due to the positive energy driving inflation. It is possible to identify those two superfields thus getting inflaton and goldstino in a single chiral superfield~\cite{kt1,kt2}. The sinflaton has to be stabilized during inflaton, with its mass beyond the Hubble value in order to get a single-field inflation favored by the Planck data. Slow-roll inflation and a de Sitter vacuum (DE) are obtained by carefully engineering the inflaton scalar potential $V_F$ in terms of a K\"ahler potential $K$ and a superpotential $W$ as $(M_{\rm Pl}=1)$ 
\be \lb{cpot}
V_F = e^K \left( |DW|^2 - 3|W|^2 \right) \quad  {\rm with} \quad DW= W' + K'W~,
\ee
where the primes denote the derivatives with respect to the chiral superfield, and then solving the so-called 
$\eta$-problem by arranging the inflationary trajectory along a flat direction of the K\"ahler potential $K$ in order to
get slow roll.  Stability of inflation (enough e-foldings) against its possible spoiling by sinflaton is a major concern in this business and is difficult to achieve. The alternative is to get rid of sinflaton by the use of nilpotent (constrained) chiral superfields~\cite{Komargodski:2009rz,Antoniadis:2014oya}. However, the nilpotent superfields are problematic at very high energies and in quantum theory, so we do not employ them for inflation.

The straightforward way of extending the modified $f(R)$ gravity model like the one in (1) to supergravity is possible by the use of the curved superspace of the so-called old-minimal supergravity that gives rise to the 
action~\cite{Ketov:2010qz}
\be \lb{msg}
S = \int d^4x d^4\theta E^{-1} N(\car,\bar{\car}) +\left[ \int d^4x d^2\Theta 2\ce F(\car) +{\rm h.c.}
\right]
\ee
in terms of the supergravity chiral superfield $\car$ having the scalar curvature $R$ amongst its field components 
at $\Theta^2$, and the two potentials $N$ and $F$.~\footnote{We use the standard notation of Wess and 
Bagger in superspace~\cite{Wess:1992cp}.} The action (\ref{msg}) can be transformed into the standard matter-coupled Einstein supergravity action with two chiral matter superfields~\cite{Cecotti:1987sa,Gates:2009hu}. However, it cannot properly embed the Starobinsky inflationary model (1)  because of the extra propagating scalars that may be ghost-like and whose potential is generically unbounded from below. Those unwanted scalars have to be stabilized either by introducing more superfields or by tuning the functions $N$ and $F$, while no good example  was ever found.

\section{Inflaton in a vector multiplet}

To solve the problems mentioned above, we propose 
\begin{itemize}
\item the use of a massive vector multiplet instead of a chiral multiplet in order to unify real inflaton scalar and goldstino (a massive vector multiplet has only one real scalar!);
\item the use of the super-Higgs-effect in order to give mass to the vector multiplet;
\item the use of the Born-Infeld (BI) kinetic term instead of the (canonical) Maxwell kinetic term for the vector multiplet 
in order to generate the goldstino action for the fermionic superpartner of the inflaton with the F-type (high scale) spontaneous SUSY breaking after inflation;
\item the use of the alternative Fayet-Iliopoulos (FI) term without gauging the R-symmetry in order to generate the additional D-type (small scale) spontaneous SUSY breaking after inflation for uplifting the Minkowski vacuum to a de Sitter vacuum (dark energy).
\end{itemize}

\def\pa{\partial}

On the one hand, it is expected that Maxwell electrodynamics does not remain unchanged up to the Planck scale because of its internal problems related to the Coulomb singularity and the unlimited values of electro-magnetic field. This motivated Born and Infeld \cite{Born:1934gh} to propose the non-linear vacuum electrodynamics with the Lagrangian (in flat spacetime)
\be \lb{binf}
{\cal L}_{\rm BI} = -M_{\rm BI}^4\sqrt{-\det\left(\h_{\m\n}+M_{\rm BI}^{-2}F_{\m\n}\right)}=-M_{\rm BI}^4
-\frac{1}{4}F^2 +{\cal O}(F^4)~,
\ee
where $\h_{\m\n}$ is Minkowski metric, $F_{\m\n}=\pa_{\m}A_{\n} -\pa_{\n}A_{\m}$, and
$F^2=F^{\m\n}F_{\m\n}$.   The BI theory has the new scale $M_{\rm BI}$ whose value cannot exceed the GUT scale where electro-magnetic interactions merge with strong and weak interactions. The BI theory naturally emerges (i) in the bosonic part of the open superstring effective action, (ii) as part of Dirac-Born-Infeld (DBI) effective action of a D3-brane, and (iii) as part of Maxwell-Goldstone action describing partial supersymmetry breaking of $N=2$ supersymmetry to $N=1$ supersymmetry. The peculiar non-linear structure of the BI theory is responsible for its electric-magnetic (Dirac) self-duality, taming the Coulomb self-energy of a point-like electric charge, and causal wave propagation (no shock waves and no superluminal propagation).

On the other hand, the universal goldstino action has the Akulov-Volkov (AV) Lagrangian in flat 
spacetime~\cite{Volkov:1973ix},
\be \lb{avl}
\mathcal{L}_{\rm AV} = -M^4_{\rm susy}\det\left(\d^a_b+\fracmm{i}{2M^4_{\rm susy}}\bar{\l}\g^a\pa_b\l\right) =  -M^4_{\rm susy} -\frac{i}{2}\bar{\l}\g\cdot\pa\l +{\cal O}(\l^4)~,
\ee
where $\l(x)$ is a Majorana fermion field of spin 1/2. This fermionic field is called the {\it goldstino} because the AV action has the spontaneously broken non-linearly realized rigid SUSY under the transformations
\be \lb{avs} 
\d\l= M^2_{\rm susy}\ve +\fracmm{i}{M^2_{\rm susy}}(\bar{\ve}\g^a\l)\pa_a\l
\ee
with the infinitesimal Majorana spinor parameter $\ve$. The AV theory (\ref{avl}) has the cosmological constant $M^4_{\rm susy}$ where $M_{\rm susy}$ is the spontaneous SUSY breaking scale. A coupling of the AV action to supergravity is supposed to generate a gravitino mass via the so-called super-Higgs effect \cite{Wess:1992cp} when the gravitino "eats up" the goldstino and thus gets the right number of the physical degrees of freedom.

The manifestly supersymmetric extension of the BI action minimally coupled to supergravity in curved superspace of the (old-minimal) supergravity is given by
\be \lb{ssbi}  
S_{\rm sBI}=  \frac{1}{4}\left( \int d^4xd^2\theta {\cal E} W^2 + {\rm h.c.}\right) +
\frac{1}{4} M_{\rm BI}^{-4} \int d^4xd^2\theta d^2\bar{\theta} E\fracmm{ W^2{\bar W}^2}{1+\frac{1}{2}A+\sqrt{1+A+\frac{1}{4}B^2}}
\ee
where $A=\frac{1}{8}M_{\rm BI}^{-4}\left( {\cal D}^2W^2+{\rm h.c.}\right)$ and 
$B=\frac{1}{8}M_{\rm BI}^{-4}\left( {\cal D}^2W^2-{\rm h.c.}\right)$. The $W^{\a}$ is the chiral  gauge-invariant field strength, $W_{\a}=-\fracm{1}{4}\left( \bar{\cal D}^2-4{\cal R}\right){\cal D}_{\a}V$,
of the gauge real scalar superfield pre-potential $V$ describing an $N=1$ vector multiplet \cite{Wess:1992cp}.

The identification of the photino $\l$ with the goldstino of the spontaneously broken local SUSY requires  
$M_{\rm BI} = M_{\rm susy}$.   The ultimate recovery of the AV action from the super-BI action is possible by identifying the goldstino $\l_{\a}$ with the leading field component of the superfield $W_{\a}$ and projecting  the other fields out, $F_{\m\n}(A)=D=\j_{\m}=0$ in the absence of gravity, $e^a_{\m}=\d^a_{\m}$. Then the super-BI action reduces to the AV action up to a field redefinition in the higher order terms (with respect to $\l$).

A generic  Lagrangian of the massive vector multiplet is governed by a real potential $J$, 
\begin{equation}
\label{hsslag}
\mathcal{L}=\int d^2\theta 2\mathcal{E}\left\lbrace \frac{3}{8}(\bar{\mathcal{D}}\bar{\mathcal{D}}-8\mathcal{R})e^{-\frac{2}{3}J}+\frac{1}{4}W^\alpha W_\alpha
\right\rbrace +{\rm h.c.}~,
\end{equation}
while its bosonic part in Einstein frame reads $(M_{\rm Pl}=1)$
\begin{equation} \label{complag}
e^{-1}\mathcal{L}=\frac{1}{2}R -\frac{1}{4}F_{mn}F^{mn}-
\frac{1}{2}J''\partial_mC\partial^mC-\frac{1}{2}J''B_mB^m-\frac{g^2}{2}{J'}^2~,
\end{equation}
where $C=\left. V\right|$ is the real scalar inflaton field and $J=J(C)$. Hence, any scalar potential given by a real function squared can be supersymmetrized~\cite{Farakos:2013cqa,Ferrara:2013rsa}. For example, the Starobinsky potential is obtained with 
the choice ($M_{\rm Pl}=1$ and $g=1$)
\begin{equation} \label{starj}
 J(C)=  \frac{3}{2} \left( C- \ln C\right)\quad {\rm and} \quad 
 C =  \exp\left( \sqrt{2/3} \phi\right)~.
 \end{equation}

The master function $J(V)$ can be generalized to a function $\tilde{J}(He^{2V}\bar{H})$  where we have introduced the Higgs chiral superfield $H$. The  $\tilde{J}$ is invariant under the gauge transformations  
\begin{equation}
 \label{sgtr}
H\to e^{-iZ}H~,\quad \bar{H}\to e^{i\bar{Z}}\bar{H}~,\quad V\to 
V+ \frac{i}{2}(Z-\bar{Z})~,
\end{equation}
whose gauge parameter Z itself is a chiral superfield. The original theory of the massive vector multiplet governed by the master function $J$  is recovered in the supersymmetric gauge $H=1$. We can also choose 
the different (Wess-Zumino) supersymmetric gauge in which $V=V_1$, where $V_1$ describes the irreducible  massless vector multiplet minimally coupled to the dynamical Higgs chiral multiplet $H$.  The standard Higgs mechanism appears when choosing the canonical function $J=\frac{1}{2} He^{2V}\bar{H}$ that corresponds to a linear function $\tilde{J}$. The whole procedure is known as the super-Higgs mechanism  \cite{Wess:1992cp}.

In the models introduced above, SUSY is always restored after inflation. To avoid it, we add Polonyi chiral superfield  (the hidden sector) with 
\begin{equation} \label{polonyi}
K= \Phi\bar{\Phi}~,\qquad \mathcal{W}=\mu(\Phi +\beta)~.
\end{equation}
This defines the so-called Polonyi-Starobinsky (PS) 
supergravity~\cite{Aldabergenov:2016dcu,Aldabergenov:2017bjt}. 

The total Lagrangian of the PS supergravity reads
\begin{equation}
\label{hsslag2}
\mathcal{L}=\int d^2\theta 2\mathcal{E}\left\lbrace \frac{3}{8}(\bar{\mathcal{D}}\bar{\mathcal{D}}-8\mathcal{R})e^{-\frac{1}{3}(K+2J)}+\frac{1}{4}W^\alpha W_\alpha +\mathcal{W}(\Phi) 
\right\rbrace +{\rm h.c.}
\end{equation}
and leads to a Minkowski vacuum after inflation, though with spontaneously broken SUSY.  However, it also leads to the the F-type contribution of Polonyi scalar to the inflationary scalar potential, due to its mixing with the inflaton, while that contribution spoils slow roll inflation. This can be cured by adding the field-dependent FI term (see below), whose coefficient is a function of the Polonyi superfield, and changing the $J$-function in (\ref{starj})
appropriately~\cite{Aldabergenov:2017hvp,Aldabergenov:2018nzd}.

The F-type SUSY breaking after inflation at {\it arbitrary} scale can be achieved by choosing the free parameter 
$\mu$ in (\ref{polonyi}). To simultaneously generate a tiny positive cosmological constant by uplifting the Minkowski vacuum to a de Sitter vacuum, we employ the extra (D-type) spontaneous SUSY breaking provided by the alternative 
FI term~\cite{Cribiori:2017laj}. The vector multiplet $V$ can be decomposed as $V={\cal V} +G$, where $G$ is the goldstino superfield, $G^2=0$. The alternative FI term is just proportional to $G$. Explicitly, it takes the form
\cite{Cribiori:2017laj,Kuzenko:2018jlz}
\begin{equation}
{\cal L}_{\text{FI1}}=2\xi \int d^4\theta E\fracmm{W^2\bar{W}^2}{\mathcal{D}^2 W^2\bar{\mathcal{D}}^2\bar{W}^2}
\quad {\rm or} \quad 
{\cal L}_{\text{FI2}} = 2\xi \int d^4\theta E\fracmm{W^2\bar{W}^2}{(\mathcal{D}W)^3}~,
\end{equation}
with the coupling constant $\xi$.  The elimination of the auxiliary field $D$ gives rise to the positive cosmological constant equal to $\frac{1}{2}\xi^2$ that can be identified with the observed value.

\section{Phenomenology of PS supergravity with FI term}

Phenomenology of the PS supergravity with the alternative FI term was also systematically studied~\cite{Addazi:2017ulg,Addazi:2018pbg}. We find that
\begin{itemize}  
\item in the early Universe gravitinos are produced from vacuum during inflation and via inflaton and Polonyi decays;
\item matching the known DM abundance with the produced gravitinos allows us to fix the gravitino mass as  $m_{3/2}\approx 7.7 \cdot 10^{12}$ GeV;
\item the F-type SUSY breaking scale is dictated by the coefficient $\mu$ at the linear term in the Polonyi superpotential and is given by $\mu^{1/2}\sim\sqrt{m_{3/2}M_{\rm Pl}}$ that must be close to $M_{\rm GUT}$;
\item  the de Sitter vacuum (the cosmological constant $\L_0$ or DE) is obtained by tuning the coefficient at the alternative 
FI term as $\frac{1}{2}\xi^2=\Lambda_0$;
\item there are no gravitino and Polonyi overproduction problems, and the Big Bang Nucleosynthesis (BBN) constraints are avoided;
\item the gravitino is LSP; all other sparticles masses are above the $m_{3/2}$ in this high-scale SUSY scenario and, therefore, are phenomenologically irrelevant;
\item the reheating temperature after inflation is about $10^{10}$ GeV.
\end{itemize}

\section{BI and FI in supergravity}

The natural question arises: is it possible to get rid of the Polonyi superfield and use merely a single vector
multiplet for all purposes of describing the Dark Side of the Universe? The answer is in affirmative~\cite{Abe:2018plc,Abe:2018rnu} though  we have to use all our tools such as the BI structure and the alternative 
FI terms.

After eliminating the auxiliary fields and Weyl rescaling to Einstein frame, $e\rightarrow e^{4{\cal J}/3}e$ and $g^{mn}\rightarrow e^{-2{\cal J}/3}g^{mn}$, the bosonic part of our supergravity model with the BI structure and the (first) FI term reads
\begin{align}
e^{-1}\mathcal{L}_{\text{I}}= & \frac{1}{2}R-\frac{1}{2}{\cal J}''\partial_aC\partial^aC-\frac{1}{2}{\cal J}''B_aB^a \nonumber \\
& +\frac{e^{4{\cal J}/3}}{8\alpha}\left[1-\sqrt{\displaystyle 1+8\alpha Z^2}\sqrt{\displaystyle 1+
4\alpha F^2e^{-4{\cal J}/3}+4\alpha^2(F\tilde{F})^2}\right]~,
\end{align}
where $Z\equiv \frac{{\cal I}}{4}-{\cal J}'e^{-2{\cal J}/3}$,  $\tilde{F}_{ab}\equiv-\frac{i}{2}\epsilon_{abcd}F^{cd}$, $B_a$ is the vector field whose field strength is $F_{ab}$. The absence of ghosts requires ${\cal J}''>0$.

In particular, the auxiliary field $D$ is eliminated via its algebraic equation of motion,
\begin{equation}
D=\fracmm{Z}{\sqrt{1-8\alpha Z^2}}\sqrt{\displaystyle 1+4\alpha F^2e^{-4{\cal J}/3}+4\alpha^2(F\tilde{F})^2}~,
\end{equation}
so it can have the non-vanishing vacuum expectation value, $\langle D\rangle\neq 0$ or $\langle Z\rangle\neq 0$, that spontaneously breaks SUSY.  

The scalar potential is given by 
\begin{equation}
\mathcal{V}=\frac{e^{4{\cal J}/3}}{8\alpha}\left(\sqrt{1+8\alpha Z^2}-1\right)~.\label{V}
\end{equation}
and can be used to realize both slow-roll and Starobinsky-like inflation, as well as a de Sitter vacuum, by properly choosing the real functions ${\cal J}$ and $Z$.

Similarly, as regards the second alternative FI term, we find the D-dependent terms in Jordan frame as follows:
\begin{align}
e^{-1}\mathcal{L}_{\text{II}}(D)=&-\fracmm{{\cal I}}{16}\left[4D-\fracmm{4F^2}{D}+\fracmm{F^4-(F\tilde{F})^2}{D^3}\right]+e^{-2{\cal J}/3}{\cal J}'D\nonumber\\&+\fracmm{1}{8\alpha}\left(1-\sqrt{1+4\alpha(F^2-2D^2)+4\alpha^2(F\tilde{F})^2}\right)~.\label{LD}
\end{align}

The generic new features are: (i) the non-polynomial dependence upon $D$,  (ii) the highly non-linear dependence upon $F$, (iii) the non-trivial coupling to chiral matter for respecting the K\"ahler gauge invariance, and (iv) 
the rich structure of the scalar potential.

\section{Conclusion}

We find that the unification of inflation, DE and DM, as well as the unification of the associated scales are possible in supergravity. The origin of inflaton, DE and DM is entirely (super)gravitational in our approach. The non-standard supergravity tools (the real $J$ potential, the BI structure of the kinetic terms, the alternative FI term) are essential. In particular, the FI term  uplifts the Minkowski vacuum in Starobinsky model to a de Sitter vacuum (DE). The high-scale SUSY  is required. The only "low-energy" impact of the high-Scale SUSY is the gravitino LSP as the DM. Therefore, accelerator searches for SUSY are expected to produce no signs of SUSY, and then only cosmoparticle physics probes of SUSY are possible~\cite{Ketov:2019mfc}.

Finally, we add a comment on the possibility of describing the Primordial Black Holes (PBHs) in supergravity. The PBHs may be formed in the early Universe by collapsing of primordial density perturbations resulting from inflation, when those perturbations reenter the horizon and are large enough (in other words, when gravity forces are larger than pressure). Apart from being considered as the alternative or additional (non-particle) source for DM, some PBHs (of stellar mass) are also considered as the candidates for the gravitational wave effects caused by the binary black hole mergers observed by LIGO/Virgo collaboration. The PBHs abundance $f=\O_{\rm PBH}/\O_{\rm cr.}$ is proportional to the amplitude of the scalar perturbations $P_{\z}$  and is also proportional to $(M_{\rm Sun}/M_{\rm PBH})^{1/2}$. Hence, for the PBH to be the DM, one needs the mass $M_{\rm PBH}\approx
10^{-12}M_{\rm Sun}$, and the enhancement of the perturbation spectrum from $10^{-9}$ (CMB) to $10^{-2}$ (PBH) at the last stages of inflation. 

In a single-field inflation, perturbations are controlled by the inflaton scalar potential, so that large fluctuations are produced when the slow roll parameter $\ve=r/16$ goes to zero, i.e. when the potential $V$ has a near-inflection point with $V'\approx V''\approx 0$. To unify a copious PBH production with the CMB observables, those events should be "decoupled" by demanding the existence of another (relatively short) plateau in the scalar potential after the inflationary plateau towards the end of inflation. This is not the case for the Starobinsky inflation but can be easily achieved in our supergravity framework by allowing another plateau or spikes of the $J'$-function in the inflaton scalar potential $V=\frac{1}{2}g^2 (J')^2$.

\section*{Acknowledgements}

The author was supported by Tokyo Metropolitan University, the World Premier International Research Center Initiative (WPI), MEXT, Japan, and  the Competitiveness Enhancement Program of Tomsk Polytechnic University in Russia.


\end{document}